# Direct observation of growth and collapse of a Bose-Einstein condensate with attractive interactions

Jordan M. Gerton, Dmitry Strekalov*, Ionut Prodan, and Randall G. Hulet

*Physics and Astronomy Department and Rice Quantum Institute, MS 61, Rice University, Houston, Texas 77251, USA*
*\* Present address: Jet Propulsion Laboratory, MS 300-123, Pasadena, California 91109, USA*

**The dynamical behavior of Bose-Einstein condensation (BEC) in a gas with attractive interactions is striking. Quantum theory predicts that BEC of a spatially homogeneous gas with attractive interactions is precluded by a conventional phase transition into either a liquid or solid[1]. When confined to a trap, however, such a condensate can form[2] provided that its occupation number does not exceed a limiting value[3,4]. The stability limit is determined by a balance between self-attraction and a repulsion arising from position-momentum uncertainty under conditions of spatial confinement. Near the stability limit, self-attraction can overwhelm the repulsion, causing the condensate to collapse[5-8]. Growth of the condensate, therefore, is punctuated by intermittent collapses[9,10], which are triggered either by macroscopic quantum tunneling or thermal fluctuation. Previous observation of growth and collapse has been hampered by the stochastic nature of these mechanisms. Here we reduce the stochasticity by controlling the initial number of condensate atoms using a two-photon transition to a diatomic molecular state. This enables us to obtain the first direct observation of the growth of a condensate with attractive interactions and its subsequent collapse.**

Condensate growth is initiated by cooling the gas below the critical temperature $T_c$ for BEC. For attractive interactions, the number of condensate atoms $N_0$ grows until the condensate collapses. During the collapse, the density rises giving a sharp increase in the rate of collisions, both elastic and inelastic. These collisions cause atoms to be ejected from the condensate in an energetic explosion. The physics which determines both the stability condition and the dynamical process of collapse of the condensate, draws some interesting comparisons to a star going supernova[11], even though the time, length, and energy scales for these two phenomena are vastly different. In the stellar case, the stability criterion is provided by a balance between the pressure due to the quantum degeneracy of electrons which make up the star and gravitational attraction. If the mass of the star exceeds the stability limit[12], the star collapses, releasing nuclear energy and triggering a violent explosion. In contrast to the stellar case, the condensate regrows after a collapse as it is fed by collisions between thermal atoms in the gas, and a series of sawtooth-like cycles of growth and collapse will continue until the gas reaches thermal equilibrium[9,10]. We previously attained evidence for this nonequilibrium dynamical behavior in $^7$Li by measuring the distribution of $N_0$ at selected times following a fast quench of the gas[13]. $N_0$ was found to be distributed between small numbers, $N_0 \approx 100$, and the maximum number of ~1250 atoms, in agreement with the growth and collapse model.

The process of making reliable measurements of such small values of $N_0$ destroys the condensate, preventing an observation of condensate dynamics in real time. Although the phase-contrast imaging technique employed here has been used for (nearly) nondestructive measurements of large condensates[14], a few incoherent photons are always scattered, and these heat the gas. As the phase-contrast signal is proportional to the number of scattered photons, achieving sufficient sensitivity to small condensates, such as those studied here, results in excessive heating and destruction of the condensate[15]. In the present work, direct observation of the dynamics is made possible by dumping the condensate, while only minimally affecting the thermal atoms. This synchronizes the growth and collapse cycles for different experiments at a particular point in time. The subsequent growth/collapse dynamics are then obtained by repeating the experiment and measuring $N_0$ at different delays following the dump pulse. In a recent experiment with essentially pure condensates of $^{85}$Rb atoms, a magnetically-tuned Feshbach resonance was used to suddenly switch the interactions from repulsive to attractive and thereby induce a collapse at a specified time[16]. In that experiment, condensates are produced with $N_0$ far greater than the stability limit, and consequently with high initial energy. In contrast, for the present experiment, the collapse occurs with $N_0$ below the maximum number via macroscopic quantum tunneling or thermal fluctuation[7-9]. Furthermore, the condensate coexists with a gas of thermal atoms, allowing the kinetics to be probed.

The apparatus used to produce BEC of $^7$Li has been described previously[15]. Permanent magnets establish an Ioffe-Pritchard type trap with a nearly spherically symmetric, harmonic trapping potential. Approximately $2.5 \times 10^8$ atoms in the F = 2, $m_F$ = 2 hyperfine sublevel of $^7$Li are directly loaded into the trap from a laser-slowed atomic beam. Following loading, the atoms are evaporatively cooled to ~400 nK with ~4 $\times 10^5$ atoms using a microwave field to selectively spin-flip and remove the most energetic atoms. Under these conditions, the gas is quantum degenerate. The microwave frequency is maintained at the end frequency for two seconds following evaporation, and is subsequently lowered during a 100 ms duration quench pulse that removes all but the coldest ~$10^5$ atoms. This leaves the gas far from thermal equilibrium, and if left to freely evolve, the condensate will alternately grow and collapse many times over a period of ~30 s[13]. Destructive phase-contrast imaging is used to determine $N_0$, the total number of atoms $N$, and their temperature $T$[15].

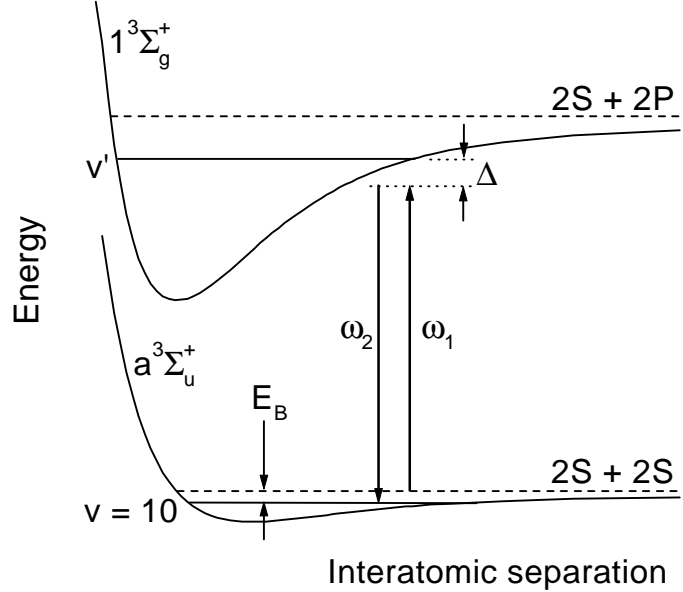

**Figure 1** Two-photon molecular association technique. Two atoms in the electronic ground state (2S + 2S) interact along the $a^3\Sigma_u^+$ diatomic molecular potential at the dissociation limit indicated by the dashed line. They may be associated into the least-bound vibrational level, v = 10, when the relative frequency between two drive lasers, $\omega_1 - \omega_2$, is equal to $E_B$, where $E_B \approx$ 12.47 GHz is the binding energy of the v = 10 level. Each laser is tuned near resonance with a vibrational level v′ = 72, of the $1^3\Sigma_g^+$ excited state potential. The intensity of each laser beam is between 0.5 and 2.5 W/cm$^2$ and the intermediate state detuning $\Delta \approx$ -110 MHz.

Following a delay of several seconds after the microwave quench pulse, the condensate is dumped by a light pulse consisting of two co-propagating laser beams whose frequency difference is tuned to resonance between the collisional state of two free atoms and a vibrational level of the diatomic molecule Li$_2$, as shown in Fig. 1. Once in the molecular state, the laser of frequency $\omega_2$ can stimulate a single-photon transition to the intermediate level v′, which may spontaneously decay, most likely into a state of two energetic atoms that will escape the trap. This method for removing atoms is very energy specific since the observed two-photon linewidth of δ500 Hz is much less than the ~5 kHz thermal energy spread of the trapped atoms. In particular, the condensate may be selectively removed without significantly affecting the remaining atoms. This is demonstrated in Fig. 2, where both the measured condensate fraction and the fraction of total atoms are plotted as a function of the duration of the light pulse. The two-photon spectroscopy of this work is similar to that performed previously in a non-quantum-degenerate gas of lithium atoms at $T \approx 1$ mK[17], and in a Bose condensate of Rb atoms[18].

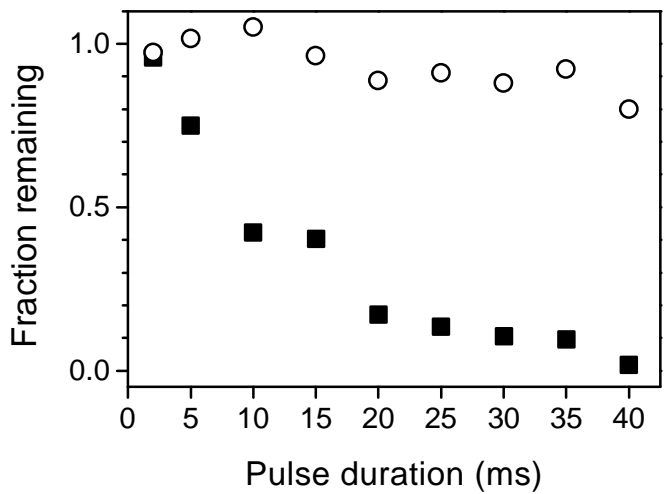

**Figure 2** Energy selectivity of the two-photon light pulse. The fraction of condensate atoms (solid squares) and total atoms (open circles) which remain trapped immediately following a light pulse is plotted as a function of the light pulse duration. Each data point is an average of five separate experiments normalized to measurements taken without the light pulse. For these measurements, $N = (7 \pm 1) \times 10^4$ atoms. Achieving the necessary energy selectivity required the two diode lasers used to drive the two-photon transition to be phase-locked. When locked, the relative frequency spectrum of the two lasers was measured to be less than 1 Hz in width.

Following the light pulse, the gas is allowed to freely evolve for a certain time, at which point a destructive measurement of $N_0$ is made. Figure 3a shows the dynamical evolution of the condensate following a light pulse whose duration is adjusted to reduce $N_0$ to an initial value of ~100 atoms. $N_0$ increases immediately following the light pulse as the condensate is fed via collisions between noncondensed thermal atoms, reaching a maximum value consistent with the expected upper limit of 1250 atoms. A collapse is clearly indicated by the subsequent reduction in $N_0$. After the collapse, $N_0$ grows again, since the gas is not yet in thermal equilibrium. Figure 4 shows representative phase-contrast images taken from the data in Fig. 3a for the specified delay times. The central peak, most clearly visible at 450 ms delay, corresponds to the condensate.

The condensate growth rate may be adjusted by varying the duration of the light pulse. By reducing the duration, fewer atoms are removed from both the condensate and from the low-energy thermal atoms that directly contribute to condensate growth, and consequently the growth rate increases. This is illustrated in Fig. 3b, where two secondary peaks are now discernible. For Fig. 3c, the light pulse duration is lengthened, causing more atoms to be removed, and slowing the rate of growth.

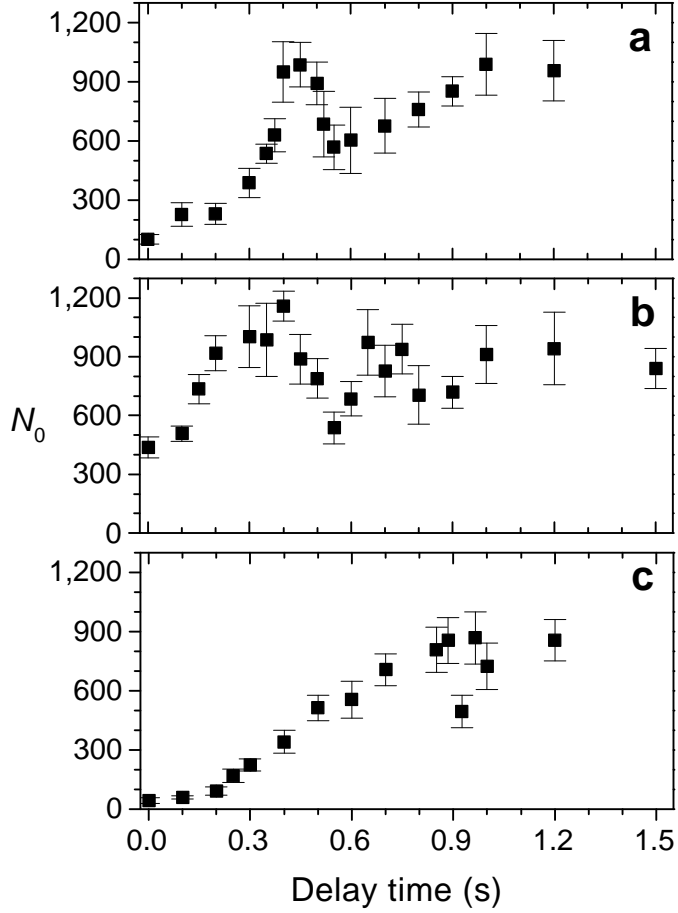

**Figure 3** Condensate growth and collapse dynamics. Each data point is the mean of 5 measurements for the same delay time following the light pulse and the error bars are the standard deviation of the mean. a: $N_0$ is reduced to ~100 atoms using the two-photon light pulse. The reduction in $N_0$ after the first peak at 450 ms is a direct manifestation of collapse. b: The condensate is only partially dumped in order to speed up the initial growth. Subsequent maxima and minima are observed as the gas continues to undergo growth and collapse cycles while evolving towards thermal equilibrium. c: The light pulse duration is increased, so that the condensate is dumped completely, to within the experimental resolution. Note the slow turn on of growth and the subsequent saturation in the growth rate (see text). For a and c, there is a 3 s delay following the microwave quench pulse before the light pulse, while in b, the delay is 5 s. Additionally, for a and c, $N = (7 \pm 1) \times 10^4$ atoms and $T = (170 \pm 15)$ nK immediately before the light pulse, while for b, $N = (1.0 \pm 0.1) \times 10^5$ atoms and $T = (200 \pm 20)$ nK. For each individual image, the statistical uncertainty in $N_0$ is ±60 atoms, while the systematic uncertainty, dominated by uncertainties in the imaging system, is ±20%[13].

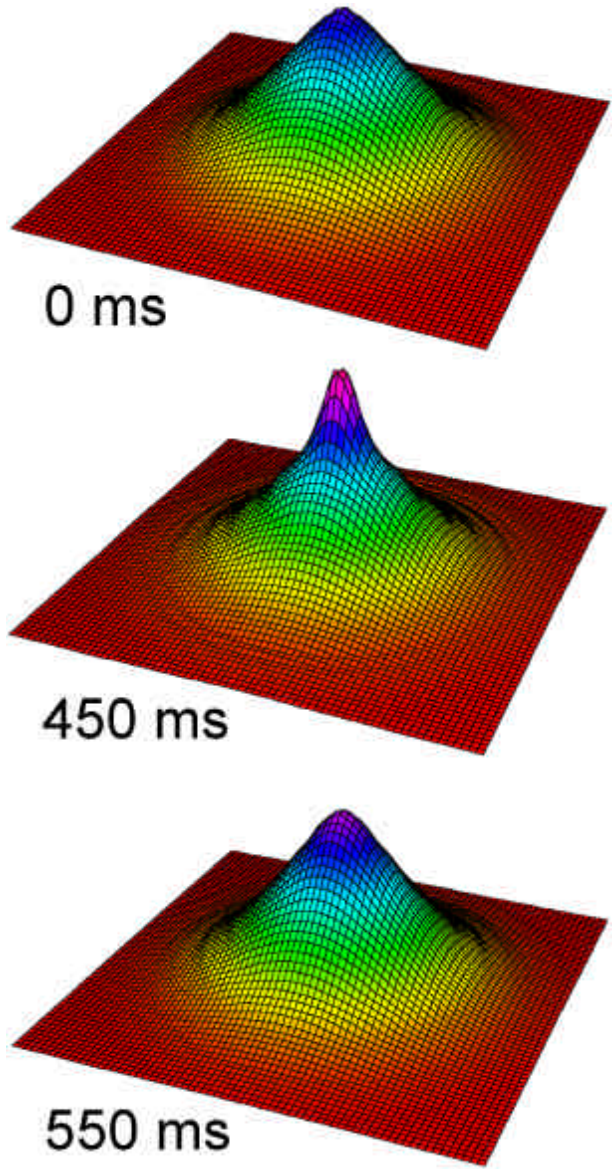

**Figure 4** Phase-contrast images. These images were selected from those used to construct Fig. 3a. The upper image corresponds to data immediately after the dump pulse (0 ms), the middle to the peak of condensate growth (450 ms), and the lower to the first collapse (550 ms). The fitted values of $N_0$ are 40, 1210 and 230 atoms for the upper, middle and lower images, respectively. The displayed images result from angle-averaging the actual images about the probe laser propagation axis and the signal height is proportional to the column density of the atom cloud integrated along this direction[4]. The image area corresponds to an 85 μm square. A movie of the growth and collapse was produced using representative time-ordered images from the data of Fig. 3a, and is available for viewing at http://atomcool.rice.edu/collapse.html.

Each data point in Fig. 3 is the mean of five separate measurements of $N_0$. Consequently, these data represent an average of many trajectories whose initial phase and rate of growth differ slightly. To analyze the results, we numerically simulate the collisional redistribution of atoms over the energy states of the trap using the quantum Boltzmann equation, as described in detail elsewhere[9]. The colored curves shown in Fig. 5 are a sample of simulated trajectories which include the effect of the microwave quench pulse and the light pulse. The variation in condensate growth following a light pulse is mainly the result of slight differences in initial conditions which lead to variations in the energy distribution of atoms in the trap. Additionally, the stochastic nature of the collapse process, which causes each collapse to occur at a slightly different value of $N_0$, contributes to dephasing of different trajectories. The heavy black line represents the average of 40 trajectories obtained by multiply running the simulation with slightly different initial conditions. In the simulations, $N_0$ is set to 100 atoms immediately following the light pulse in order to provide direct comparison with the data in Fig. 3a. Additionally, $N_0$ is set to 200 atoms immediately following a collapse in order to achieve the best agreement with our previous statistical studies[13]. The only adjustable parameter in the simulations was the fraction of thermal atoms lost within the spectral width of the two-photon transition.

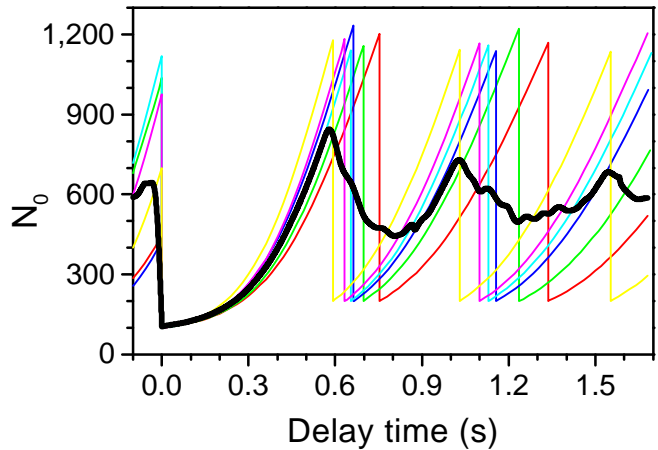

**Figure 5** Simulation of condensate dynamics. The colored curves show individual trajectories generated by a numerical quantum Boltzmann simulation, which models the kinetics of evaporative cooling and thermal equilibration. Individual trajectories were generated by running the simulation for 40 different initial values of $N$ spaced evenly between $2.4 \times 10^8$ and $2.6 \times 10^8$ atoms. The resulting values of $N$ just before the light pulse range between $6 \times 10^4$ and $8 \times 10^4$ atoms, in agreement with the range of $N$ observed in the data. The heavy solid line is the mean of these 40 trajectories. The effect of the light pulse on the thermal atoms is simulated by instantaneously reducing the population of atoms within a 500 Hz wide Lorentzian energy band, consistent with the observed spectral width. For the simulation shown, 90% of the thermal atoms at the Lorentzian peak are lost.

The simulation results agree well with the data in several respects. The data in Fig. 3b show that each subsequent peak following the initial growth is slightly lower, as the trajectories corresponding to each individual measurement dephase from one another. This dephasing causes the maxima (minima) to occur at smaller (larger) values of $N_0$ than for any individual trajectory. This behavior is seen in the simulation average shown in Fig. 5, confirming our understanding of the role of averaging in these measurements.

Condensate growth should be affected by the quantum statistical effect known as Bose enhancement: the condensate growth rate scales with the occupation number $N_0$. This would lead to an exponentially increasing growth rate, which is, however, neither observed in the data nor predicted by the quantum Boltzmann equation model. For the conditions shown in Fig. 3c, for example, both the model and the data show a saturation in the growth rate when $N_0$ becomes larger than ~150 atoms. An analysis of the distribution of atoms among the trap energy levels in the model simulations indicates that the condensate is fed mostly by collisions between atoms with the lowest energies. Further, this analysis indicates that when the low-energy population is depleted, the growth is limited by the rate for non-Bose-enhanced 'trickle-down' collisions between high-energy atoms. Due to our high sensitivity to small values of $N_0$, the very early stages of condensate growth following the dump can be observed, illuminating this subtle, yet important, departure from the pure Bose enhancement prediction. The work of Miesner et al.[19] was the first observation of condensate growth, but that experiment could not detect small values of $N_0$ and therefore was not sensitive to the initial stages of growth.

The data presented here are the first direct observations of the growth and collapse of Bose-Einstein condensates with attractive interactions. By enabling the preparation of condensates with a desired mean number of atoms with minimal perturbation to the noncondensed atoms, the two-photon technique provides a new way of studying the dynamics of this non-equilibrium and stochastic quantum system.

**Acknowledgements.** We thank Cass Sackett for help with the quantum Boltzmann simulation. This work was supported by the U.S. National Science Foundation, the National Aeronautics and Space Administration, the Office of Naval Research, and the Welch Foundation.